# The Historical Perspective of Botnet Tools

Maxwell Scale Uwadia Osagie[1*], Osatohanmwen Enagbonma[1] and Amanda Iriagbonse Inyang[1]

[1]*Department of Physical Sciences, Faculty of Science, Benson Idahosa University, P.M.B 1100, GRA, Benin City, Edo State, Nigeria.*

*Authors' contributions*

*This work was carried out in collaboration between all authors. Author MSUO designed the study, performed the statistical analysis, wrote the protocol, and wrote the first draft of the manuscript. Authors OE and AII managed the analyses of the study. All authors managed the literature searches, read and approved the final manuscript.*

## ABSTRACT

Bot as it is popularly called is an inherent attributes of botnet tool. Botnet is a group of malicious tools acting as an entity. Furthermore, history has it that the aim of what gave rise to botnet was the idea to simplify the method of message exchange within networking platform. However, this has led to several botnet tools ravaging the server environments in recent times. The working principle of these botnet tools is to get client systems that are vulnerable and thereafter, steal valuable credentials. This work is part of a comprehensive research work into botnet detection mechanism but, on this paper it primarily look at how botnet as threat tool began, the trend since inception and as well as few approaches that have been used to curb it.

_________________________________________________________________________________

*\*Corresponding author: E-mail: mosagie@biu.edu.ng;*

## 1. INTRODUCTION

Botnet as a threat has it inherent attributes to the traditional way of insecurity. Security traditionally is dynamic, every nation has tried and is still trying to ensure law and order prevail in the manner at which things are done. But, the divergent views in cultures and religion of a member state made insecurity not only a style but an act in human daily activity. In the traditional security which is the securing of humans from hazardous attack, it is currently being seen as an offspring of a modern day societal quest for power and its prevailing challenges are eating deeper by the day. Internet is a borderless network that allows different participants expression of interest which could be seen as legitimate to a particular country and view as crime in other countries. Irrespective of how this is seen, the fundamental problem of what crime represents is seen from the negative feedback on the generality of peace [1].

The Federal Networking Council (FRC) in October 1995 agreed that internet shall be defined as information system that is logically linked globally with a distinct address space that conform with the internet protocol (IP) or a subnet system that has ability to support communication with the transmission control protocol/internet protocol (TCP/IP) suits as well as an embedded platform of supporting and synchronizing segmented infrastructure. The purpose for this agreement was to guarantee access to information no matter where it is stalled. History has it that in 1960s, there was packet switching introduction, same 1960s ARPANET was developed which was directly sponsored by ARPA and having nodes such as SRI, UCLA UCSB, U.Utah. The TCP/IP came in to existence in the proposal of Cerf/Kahn in 1974 and in the year 1980 1PV4 was introduced to become ARPANET and as well accepted the TCP/IP suits. All these, metamorphose into internet becoming business/commercial entity in 1995 [2,3].

The commercial viability of internet since the transformation had made people embraced its sustainability. People sees internet as a commercial destination in 21$^{st}$ century. Hence, creating thousands of jobs and numerous opportunities to its end users as well as leading to diversification of the various computing components. This hitherto, has increased the scope of the intended idea of the computer and internet landscape as well bringing about smooth running of the internet world. This brings onboard the legitimate job that further offers opportunities to the growing entrepreneurs who over the years have used this platform in creating wealth and rendering of community services. This era (internet surfing) saw the circumventing of the legitimate process and method. The criminal activity within the computing space has further paralyzed the trust that came with it at first introduction and they have succeeded compromising the trend associated with the internet.

This work is basically organized into 4 sections with section 3 having 2 subsections. Section 1 is the introduction. This section explained the botnet historical path background and all aspects that made internet user friendly. Section 2 is the review of the advent of botnet tools. Section 3 which is the botnet tools and its effect exposed the meaning of botnet by definition as well as the associated impact. Section 3.1 itemized the botnet tool evolution with the corresponding detection method used and Section 3.2 unveils the recent curbing mechanism of botnet tool as well as future expectations. The section 4 is the concluding part the work.

## 2. RELATED LITERTURES

Botnet uses internet relay chat (IRC) protocol and it is a mechanism for communication online. Jarkko Oikarinen in 1988 developed IRC protocol in Finland. There was patronage in the late 80s through 90s in networking and data communication [2]. Jarkko Oikarinen an ICT expert from the University Oulu, Finland, created IRC to replace the MultiUser Talk (MUT) program on the University OULUBOX [2]. The idea of IRC has been compromised because hackers' sporadically and perpetually attack client computers using IRC protocol and server. The IRC is the stream for multiple clients' communication [2,4].

From the inception of creation, Internet relay chat (IRC) was used to connect different chat rooms with the basic idea of exchanging messages and following the robustness, it gained popularity. Though, still in use but recent trend in message exchange has made it metamorphous and this is due to protocol such as "I Seek You" (ICQ), Instant Messenger Protocol (AIM), and MSN messenger now being use in networking [4]. What made these protocols better is it orientated background in Open System for Communication in Real Time (OSCAR). Botnet tools scan systems with little security control (vulnerable systems) and a compromise server made it easier for clients systems to be bot thereby making the botmaster in control. The command and control channel remains the key strategy of a botnet tool and has the ability to execute over 200 commands at a time. Botnet as a trend in computing is a serious threat [2,4].

The internet relay chat (IRC) has revolutionized lately and this is traceable to rapid growth in information and communication technologies. The IRC technology has continued to receive robust network computing. ICT has also helped network criminals creates systematic approach in botnet and this is seen from the movement of bot from local to Peer2Peer then, the ubiquitous Hypertext Transfer Protocol (HTTP) and Spy eye [5,6,7,8]. The similarity to this evolution is the fact that they target a particular server that has multiple client computers [9].

The idea behind bot-master method is to use botnet tool via server end in making client systems Zombies (bot) and makes the systems on the scale of command and control. The activity of a botnet tool within server or network can lead to service denial (DoS) by connected systems and using the affected server, it creates room for extracting useful information from the affected (zombie) systems [9,10].

## 3. BOTNET TOOLS AND ITS EFFECT

Before looking at botnet tools and the effect to client-server systems a distinctive clarity must be made between the 80s computing gadgets and the modern gadgets. Furthermore, botnet according to [11] is a pool of compromised host controlled by a bot-master. For clarity, botnet is a holistic group of threats acting as an entity on the will of the botmaster's command and control.

In a work published by [12] they reported that [11], on a report by CipherTrust that the propagation of botnet in either ways amount to 172,000 bots and later revealed that this is on a daily recruitment. From this statistic over 60 million new bots are released yearly.

Computing gadgets in recent times changed into miniature with large amount of memory and processing speed. Though, 80s shows a significant improvement from the then computing gadgets but could not be compare to modern computing gadgets. However, credit is given to 80s because the shift from $4^{th}$ to $5^{th}$ generation of computers started in this era. But, what has made today computing gadget more sophisticated is the nanotechnology driving approach.

### 3.1 The Evolution of Botnet Tool

Since the introduction of botnet in late 80s and early 1990s it has experienced robust patronage in botmasters end and other well grounded hackers who specialized in the day to today running of networking infrastructure like client-server systems. The idea behind IRC which led to problem within chat room for personal identification exploitations has become what the entire cyber world is battling to curb. The movement started as a simple IRC to numerous bots ravaging the landscape of networks today. The table below shows the classification of the evolution of botnet tool and how it has transform from just simple Eggdrop, a botnet tool kit in 1993 to the now more sophisticated IceIX in 2015.

The classification in Table 1 is the evolution of botnet tool from 1993 to 2013 as covered by [10]. 2014 and 2015 is deeply covered by Fig. 1. The table gave explicit outlook to the year the various botnets came into network domain and with detail insight to the proliferation within speculated period. The bots itemized in the table above have one similar goal and that is taking over networking environment through command and control execution. Another thing that seems similar is the navigation processes of all botnet in different architectures.

Eggdrop as shown in Table 1 was released in 1993 without a corresponding estimated of it penetration as reported by Wang (2003) but it was the gateway to other tools such as GTbot which was mIRC as published by Janssen 2011. Confick, Torpig, and NetBus were new dimension into the different spectrum of botnet as tool. The malwares or Trojan were shift from the popularly know IRC. Hackers found out that bot could gain access to server through other protocols and this mark the beginning of network exploration. The details as stated in Table 1 have references detailing when and how they were published and for clarification see the various reference point as itemized.

Fig. 1 is a secondary data chart unveiling the new set of bots from 2014 to 2015 as it is explicitly captured bellow. From the chart, it is understood that the deepest penetration of this banking bot are the Bugot with 2250, KING with 2150, Gozi with 3450, Dyre with 2058 etc. This statistical impart of botnet made available by Counter Terrorist Unit (CTU) researchers in 2016 has further exposed to the end users the improving activities of botmasters using different botnet tools.

This impact will continue because the botnet and the botmaster approach to invading networking platform and stealing valuables via vulnerable systems are yet to be solved. Another key issue with networking threats like botnet tools is that it requires little or no knowledge to operate. Some of this tool has not gain wider publication amongst researchers. The botmaster business within networking environment is as dangerous as Trojan. Research of greater magnitude have of recent times focus on how to cube this increasing and ever rising trend that has made networking environment dreadful [13,14,15,16].

### 3.2 The Botnet and Recent Curbing Mechanism

Several mechanisms have been proposed to mitigating the risk posed by botnet tool [17]. 2018 saw a new fight back defensive mechanism called Encapsulated Detection Mechanism (EDM) [12]. is one of the most recent mechanisms design to detect and fight back botnet in a brutal force. The approach used by the EDM is an embedded structure of multiple dynamics of BSTORM algorithm and OutlieR model using One Dimensional Window Movement (ODWM) within the server network to identifying data that failed to conform to the behavioral pattern of the embedded data stream. 2018 also saw a related work, called BotDad techniques. The technique is design to identify DNS anomaly machine in an enterprise level so as to detect bot infected machine using DNS fingerprinting [18]. In 2016, [19] introduced a Bot-Meter: Charting DGA-Botnet Landscapes in Large Networks that scan through group activities within a network for possible bot. However, before the emergence of some sophisticated techniques as discussed above, several techniques as itemized in Table 1 have been used to curb botnet. The increase in the ravaging effect of botnet between 2007 and 2008 as published by Miller in 2008 was as a rising case of botnet tool called Rustock which was used in cracking down and gaining access to computers unauthorized. The malware was released in 2006 and with the intervention of the FBI a method called Operation Bot roast was introduced to help fish out the people behind the attack that was running into millions. However, the operation bot roast mainly focused its operation on IRC. For every other botnet tool itemized in the Table 1 there is a corresponding mechanism design to curb it as well as the publisher.

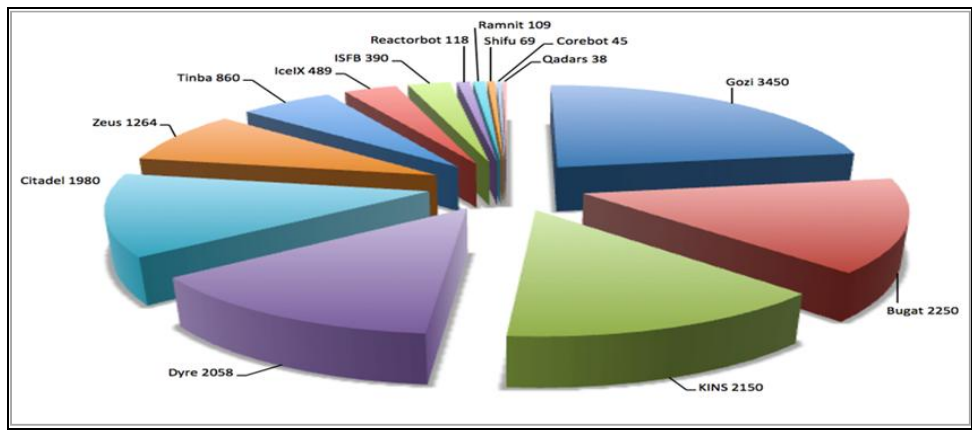

**Fig. 1. Prevalence of banking botnets in 2015 based on samples analyzed by CTU researchers [15]**

Table 1. Evolution of botnet tool

| Yr | Name | Number of estimated bot | Spam capacity (billion/d) | Aliases | Detection Approach | Type | Reference |
|---|---|---|---|---|---|---|---|
| 1993 | Eggdrop | -- | -- | Valis | -- | IRC | Wang (2003) |
| 1998 | GTBot | -- | -- | Aristotles | -- | mIRC | Janssen (2011) |
|  | NetBus | -- | -- | NetPrank | AV Software | HTTP | Wikipedia (1998) |
| 1999 | !A | 1 Billion | -- | -- | -- | -- | Wikipedia (2013b) |
| 2002 | Sdbot/Rbbot | -- | -- | IRC-SDBot | Data mining, SVM | IRC | Sevcenco (2012) |
|  | Agobot | -- | -- | W32.HLLW.Gaobot,Gaobot | Expert system | IRC | Podrezov (2013) |
| 2003 | Spybot | -- | -- | -- | -- | P2P, IRC | Schiller and Binkley (2007) |
|  | Sinit | -- | -- | Win32.Sinit, Troj/BDSinit | Network flow Analysis | P2P | Wang *et al.* (2007) |
| 2004 | Bobax | 100,000 | 27 | -- | -- | - | Kassner (2003) |
|  | Bagle | 230,000 | 5.7 | Beagle, Mitglieder | Symantec | SMTP | Symantic (2010) |
| 2006 | Rustock | 150 000 | 30 | RKRustok, Costrat | Operation b107 | IRC | Miller (2008) |
| 2007 | Akbot | 1 300 000 | -- | -- | Operation: bot roast | IRC | The H Security (2007) |
|  | Cutwail | 1 500 000 | 74 | Pandex, Mutant | -- | SMTP | Marry (2010) |
|  | Srizbi | 450 000 | 60 | Cbeplay, Exchanger | Symantec | IRC | BBC (2008) |
|  | Storm | 160 000 | 3 | Nuwar, Peacomm, Zhelatin | Fast flux | P2P | Francia (2007) |
| 2008 | Conficker | 10 500000+ | 10 | DownAndUp, Kido | AV software | HTTP/P2P | Schmudlach (2009) |
|  | Mariposa | 12 000 000 | -- | -- | Manual | IRC/HTTP | McMillan (2010) |
|  | Sality | 1 000 000 | -- | -- | Manual | P2P | Falliere (2011) |
|  | Asprox | 15 000 | -- | Sector, Kuku, Kookoo | Symantec | HTTP | Goodin (2008) |
|  | Gumblar | n/a | -- | Danmec, Hydraflux | Manual | HTTP | Mills (2009) |
|  | Waledac | 80 000 | 1.5 | -- | Kaspersky | SMTP/P2P | Goodin (2010) |
|  | Onewordsub | 40 000 | 1.8 | Waled, Waledpak N/A | -- | SMTP | Keizer (2008) |
|  | Xarvester | 10 000 | 0.15 | Rlsloup, Pixoliz | McAfee | SMTP | Symantic (2010) |
|  | Mega-D | 509 000 | 10 | Ozdok | Manual | HTTP | Warner (2010) |
|  | Torpig | 180 000 | -- | Sinowal, Anserin) | ESET | HTTP/IRC | Miller (2009) |
|  | Bobax | 185 000, | 9 | Bobic, Oderoor, Cotmonger | Manual/ BitDefender | HTTP | Symantic (2010) |
|  | Lethic | 260000 | 2 | None | Symantec | IRC | Symantic (2010) |
|  | Kraken | 495 000 | 9 | Kracken | Scan IP addresses | IRC | Jackson (2008) |
| 2009 | Maazben | 50 000 | 0.5 | -- | -- | SMTP | Symantic (2010) |
|  | Grum | 560 000 | 39.9 | Tedroo | FireEye researchers | SMTP | Danchev (2009) |

| Yr | Name | Number of estimated bot | Spam capacity (billion/d) | Aliases | Detection Approach | Type | Reference |
|---|---|---|---|---|---|---|---|
|  | Festi | n/a | 2.25 | Spamnost | ESET | SMTP/DoS | Morrison (2012) |
|  | BredoLab | 30 000 000 | 3.6 | Oficla | Symantec | HTTP/SMTP | Crowfoot (2012) |
|  | Donbot | 125 000 | 0.8 | Bachsoy | Symantec | HTTP | Stewart (2009) |
|  | Wopla | 20 000 | 0.6 | Pokier, Slogger, | Manual/PC tools | HTTP | Keizer (2008) |
|  | Zeus | 3 600 000 | n/a | Zbot, PRG,Wsnpoem | -- | -- | Messmer (2009) |
| 2010 | Kelihos | 300 000+ | 4 | Hlux | Kaspersky | P2P | Stefan (2013) |
|  | TDL4 | 4 500 000 | n/a | TDSS, Alureon | Kaspersky's TDSS killer | IRC | Kespersky (2011) |
|  | LowSec | 11 000+ | 0.5 | LowSecurity, FreeMoney | Symantec | HTTP | Symantic (2010) |
|  | Gheg | 30 000 | 0.24 | Tofsee, Mondera | Manual | DoS | Symantic (2010) |
| 2011 | Flashback | 600 000 | n/a | BacDoor.Flashback.39 | Java program | P2P | Musil (2012) |
| 2012 | Chameleon | 120 000 | -- | -- | -- | HTTP | Spider (2013) |
| 2013 | Boatnet | 500+ server computers | 0.01 | YOLOBotnet | -- | -- | Wikipedia (2013b) |

[10]

## 4. CONCLUSION

Botnet is not just a threat but a group of threats acting as an entity. The goal of the botnet is to make the activities of the network users uncoordinated and uncontrollable. Table 1 shows a transition from one botnet tool as well as the detection mechanisms designed to curbing it. However, crime and it possible solution is directly proportional because the activities created by criminals are the likely solution to stopping them if critically analyzed. As researchers continue to unveil the botnet trend and its mode of propagation within networking platforms the botmaster would continue to create different techniques meant to surpass the earlier botnet tool version used. The statistic as demonstrated by Table 1 and Fig. 1. is a clear indication of the botmaster guest for upgrade. The statistic shows a transition trend of almost a yearly improvement on different botnet tool. This paper has brought to the awareness of the general public the historical perspective of botnet tool, that is, how it began from the early stage to the recent times. Though, this is not the focus of this paper but a collaborative effort in cubing this trend with a hybrid mechanism such as EDM proposed by [12] as well as other mechanisms such as [18] will be a welcome development to putting a stop to the proliferation of botnet tool on a yearly basis.

## COMPETING INTERESTS

Authors have declared that no competing interests exist.